\pdfoutput=1
\documentclass{article}
\usepackage[utf8]{inputenc} 
\usepackage[T1]{fontenc}
\usepackage{hyperref}
\usepackage{url}      
\usepackage{setspace}
\usepackage[pdftex]{graphicx}
\usepackage{dcolumn}
\usepackage[margin=1in]{geometry}
\usepackage{amsmath}
\usepackage{graphicx}
\usepackage{float}
\usepackage{tabularx}
\usepackage[mathscr]{eucal}

\usepackage{caption}
\usepackage{subcaption}
\usepackage{multirow}
\usepackage{booktabs}

\begin{document}

\title{Warm Inflation in $f(R,T)$ gravity}

\author{ Biswajit Deb \footnote{Email: biswajitdeb55@gmail.com}\hspace{0.2cm}\textsuperscript{\textsection}, Sabina Yeasmin\textsuperscript{\textsection}, Atri Deshamukhya \\
Department of Physics, Assam University, Silchar, India \\
}

\date{}
\maketitle
\begingroup\renewcommand\thefootnote{\textsection}
\footnotetext{These authors contributed equally.}
\endgroup

\begin{abstract}
In this work, we explored warm inflation in the background of $f(R,T)$ gravity in the strong dissipation regime. Considering scalar field for FLRW universe, we derived modified field equations. We then deduced slow-roll parameters under slow-roll approximations followed by power spectrum for scalar and tensor perturbations and their corresponding spectral indices. We have considered Chaotic and Natural potentials and estimated scalar spectral index and tensor-to-scalar ratio for constant as well as variable dissipation factor $\Gamma$. We found that both the rejected potentials can be revived under the context of $f(R,T)$ gravity with suitable choice of the model parameters. Further, it is seen that within the warm inflationary scenario both the potentials are consistent with Planck 2018 bounds at the Planckian and sub Planckian energy scales.  
\end{abstract}


\section{Introduction}
The hypothesis of cosmic inflation in the early universe successfully solves the problems associated with standard big bang model like horizon and flatness problem etc.\cite{r1} The density fluctuations during inflation are believed to provide the seeds for the formation of large scale structure of the universe. The current observations from cosmic microwave background radiation (CMBR), large scale structure (LSS), WMAP and Planck support the hypothesis of cosmic inflation.\cite{r2, r3, r4, r5} In standard picture of inflation, a scalar field called inflaton is assumed to roll down its potential giving rise to the desired amount of inflation.\cite{r1, r6} Further, this field is assumed to have no interection with other fields during inflation and hence there was no scope for radiation and particle production. As a result, at the end of inflation, the universe enters a thermodynamically super cooled state which is not acceptable in big bang model where the universe is characterised by radiation and matter dominated stages.\cite{r7} To get the universe out of this cold state and putting it into the radiation dominated state was a key issue called gracefull exit problem in inflation.\cite{r1, r7, r8} To solve this problem, the concept of new inflation was introduced in which it was assumed to have a stage of reheating at the end of inflationary phase.\cite{r9} This scenario of inflationary phase followed by a reheating phase is termed as cold inflation.\cite{r7}\\

Warm inflation, an alternative theory for cold inflation was proposed by Berera in 1995 in which there is no separate reheating phase rather it is assumed that radiation production occurs simultaneously with the inflationary expansion.\cite{r10} The vacuum energy dissipates into radiation energy and hence a dissipation co-efficient $\Gamma$ is added to the Hubble damping term in the Klein-Gordon equation. During inflation, vacuum energy dominates and at the end of inflation, there is a smooth transition to radiation dominated hot big bang regime.\cite{r7, r10} \\

Apart from early inflation, it is now evident from the Type Ia Supernovae data that the universe is undergoing a second phase of accelerated expansion.\cite{r11} The reason behind this expansion is belied to be dark energy, an exotic kind of fluid with negative pressure. General Relativity (GR) can't explain this dark sector of the universe. To overcome this problem, modifications of GR is looked into and several modified theories of gravity like $f(R)$, $f(\mathscr{T})$, $f(G)$, $f(R,G)$ etc. emerged in literature where $R$ is Ricci scalar, $\mathscr{T}$ is Torsion scalar and $G$ is Gauss-Bonnet scalar.\cite{r12, r13, r14, r15} Likewise $f(R,T)$ theory of gravity was proposed by Harko {\it et al.} in 2011 where $R$ is the Ricci scalar and $T$ is the trace of the energy momentum tensor.\cite{r16} $f(R,T)$ gravity gained immediate attention as it is found to produce excellent results in the study of blackhole,\cite{r17} wormhole,\cite{r18, r19, r20, r21, r22} white dwarf,\cite{r23} pulsars,\cite{r24, r25} dark energy,\cite{r26, r27, r28, r29, r30} dark matter,\cite{r31} gravitational waves,\cite{r32, r33} bouncing cosmology,\cite{r34} scalar-tensor models,\cite{r35} anisotropic models,\cite{r36} baryogenesis\cite{r37} etc. Inflation has also been studied in $f(R,T)$ gravity and it is found that the presence of correction trace term in the theory can save many inflationary potentials those were rejected by the observational bounds.\cite{r38, r39, r40} This motivates to consider $f(R,T)$ gravity as an important tool for the study of inflationary cosmology. \\

To the best of our knowledge, warm inflation has not been studied in the frame work of $f(R,T)$ gravity till now. It will be interesting to see how $f(R,T)$ gravity affects the warm inflation scenario in the strong dissipative regime. Now, Planck 2018 data rejects many important potentials like Chaotic potential, Natural potential etc. as they fail to meet desired observational bounds on spectral index and tensor-to-scalar ratio.\cite{r5} In $f(R,T)$ gravity also, Chaotic and Natural potential fail to match Planck 2018 bounds.\cite{r38} However, Chen {\it et al.} studied the non minimal coupling of R and T in $f(R, T)$ gravity and showed that the presence of mixing term RT in the theory can rescue both Chaotic and Natural potential from rejection.\cite{r40} This provides a valid point to study whether $f(R, T)$ gravity can rescue Chaotic and Natural potential in the warm inflation scenario. \\

The paper has been organised as follows: In section 2, we review standard warm inflation and in section 3 we presented brief introduction to f(R,T) gravity in the FLRW background. In section 4, we studied warm inflation in f(R,T) gravity. In section 5, we studied warm inflation with Chaotic and Natural potentials for constant as well as variable dissipation factor. In section 6, we present our conclusion. Here, we have used natural system of unit with $c= \hbar  =1$ and the (-,+,+,+) sign convention for the metric tensor.


\section{Warm Inflation in General Relativity}

 To study warm inflation in general relativity, one starts with the following action
 \begin{equation}
     S_{GR}=\int \left(\frac{R}{8 \pi G}+L_m\right)\sqrt{-g}d^4 x
     \label{73}
 \end{equation}
Where $R$ is the Ricci scalar, $L_m$ is the matter Lagrangian density, g is the determinant of the metric tensor and G is the Newtonian gravitational constant. By applying the action principle, the Einstein field equation is given by,
\begin{equation}
    R_{\alpha\beta} - \frac{1}{2}g_{\alpha\beta}R = 8 \pi G T_{\alpha\beta}
    \label{74}
\end{equation}
where $T_{\alpha\beta}$ is the energy-momentum tensor defined as,
\begin{equation}
    T_{\alpha\beta}=-\frac{2}{\sqrt{-g}}\frac{\delta(\sqrt{-g}L_m)}{\delta g^{\alpha \beta}}=g_{\alpha \beta} L_m-2\frac{\delta L_m}{\delta g^{\alpha \beta}}
    \label{75}
\end{equation}
The Lagrangian density of the spatially homogeneous scalar field called inflaton, has the form,
\begin{equation}
    L_m^{(\phi)}=-\frac{1}{2}g^{\alpha \beta}\partial_\alpha \phi \partial_\beta \phi-V(\phi)=\frac{1}{2}\dot{\phi}^2-V(\phi)
    \label{76}
\end{equation}
where $V(\phi)$ is the potential.

The Friedmann equation and the equation of motion of inflaton field in warm inflation are
\begin{equation}
    H^2=\frac{8\pi G}{3}V(\phi)
    \label{77}
\end{equation}
 \begin{equation}
     \ddot \phi + 3H\dot \phi (1+r) + \frac{dV}{d\phi}= 0
     \label{78}
 \end{equation}
where $r=\frac{\Gamma}{3H}$ is called the dissipation rate. When $r>1$, it corresponds to  strong dissipation regime and $r<1$ corresponds to weak dissipation regime.\cite{r10} Slow-roll parameters for warm inflation are given by
 \begin{eqnarray}
\epsilon=-\frac{\dot{H}}{H^2} =\frac{M_P^2}{2(1+r)} \left(\frac{V^\prime}{V}\right)^2   
\label{79}
\end{eqnarray}
  
 \begin{eqnarray}
 \eta =-\frac{\ddot{H}}{2H\dot{H}}=\frac{M_P^2}{(1+r)} \left(\frac{V^{\prime\prime}}{V}\right) 
 \label{80}
 \end{eqnarray}
 
 \begin{eqnarray}
 \beta=-\frac{\dot{\rho}_\gamma}{H\rho_\gamma}=\frac{M_P^2}{(1+r)} \left(\frac{\Gamma^\prime V^\prime}{\Gamma V}\right)
\label{81}
 \end{eqnarray}
 The slow-roll conditions for inflationary phase are
  $\epsilon<<1$, $|\eta|<<1$ and $|\beta|<<1$.\cite{r7}
 The spectral index $n_s$ and tensor to scalar ratio $R$ are defined as \cite{r6, r41, r42, r43}
\begin{equation}
  n_s-1=\frac{d \ln P_R}{d \ln k}
  \label{82}
\end{equation}
\begin{equation}
    R=\frac{ P_T}{P_R}
    \label{83}
\end{equation} 
 where $P_R=\frac{H^3 \tau}{2\pi^2 \dot{\phi}^2}\sqrt{3(1+r)}$ is the scalar power spectrum and $P_T=\frac{16 H^2}{\pi M_P^2}$ is the tensor power spectrum, $\tau$ is the temperature of the thermal bath. (For review one can see Ref. \cite{r7})
 

\section{Dynamics of $f(R,T)$ gravity in FLRW background}
The action in $f(R,T)$ gravity theory as proposed by Harko is given by,\cite{r16}

\begin{equation}
    S = \int \left[\frac{f(R,T)}{16 \pi G} + L_m  \right] \sqrt{-g} d^4x 
    \label{a}
\end{equation}
 where $f(R,T)$ is an arbitrary function of Ricci scalar R and the trace of the energy-momentum tensor $T_{\alpha\beta}$, $L_m$ is the matter Lagrangian density, g is the metric determinant and G is the Newtonian gravitational constant. On variation of the action with respect to the metric, the $f(R,T)$ gravity field equations are obtained as,
 \begin{equation}
    f_R (R,T) R_{\alpha\beta} - \frac{1}{2} g_{\alpha\beta}f(R,T) + [g_{\alpha\beta} \nabla_\sigma \nabla^\sigma - \nabla_\alpha \nabla_\beta] f_R (R,T) =  8\pi G T_{\alpha\beta} - f_T (R,T) (T_{\alpha\beta} + \Theta_{\alpha\beta})
    \label{b}
\end{equation}
where we have denoted $f_R (R,T)= \frac{\partial f(R,T)}{\partial R}$ , $f_T (R,T)= \frac{\partial f(R,T)}{\partial T}$ and defined $T_{\alpha\beta}$ and $\Theta_{\alpha\beta}$ as,
\begin{equation}
     T_{\alpha\beta} = g_{\alpha\beta}L_m - 2 \frac{\delta L_m}{\delta g^{\alpha\beta}}
     \label{c}
\end{equation}
\begin{equation}
    \Theta_{\alpha\beta}= g^{\mu\nu} \frac{\delta T_{\mu\nu}}{\delta g^{\alpha\beta}} = -2 T_{\alpha\beta} + g_{\alpha\beta}L_m - 2 \frac{\delta^2 L_m}{\delta g^{\alpha\beta} \delta g^{\mu\nu}}
    \label{d}
\end{equation}
This term $\Theta_{\alpha\beta}$ plays a crucial role in $f(R,T)$ gravity. Since it contains matter Lagrangian $L_m$, depending on the nature of the matter field, the field equation for $f(R,T)$ gravity will be different. \\ \\
Now, in this work we have considered the simplest form of $f(R,T)$ which is $f(R,T) = R + 16 \pi G \lambda T$, where $\lambda$ is the model parameter. With this particular $f(R,T)$ form, the action becomes,
\begin{equation}
    S = \int \left[\frac{R}{16 \pi G} +  \lambda T + L_m \right] \sqrt{-g} d^4x
    \label{e}
\end{equation}
and the field equation takes the following form,
\begin{equation}
    R_{\alpha\beta} - \frac{1}{2}g_{\alpha\beta}R = 8 \pi G T_{\alpha\beta}^{(eff)}
    \label{f}
\end{equation}
where $T_{\alpha\beta}^{(eff)}$ is the effective stress-energy tensor given by,
\begin{equation}
    T_{\alpha\beta}^{(eff)}= T_{\alpha\beta} - 2 \lambda( T_{\alpha\beta} - \frac{1}{2}T g _{\alpha\beta} + \Theta_{\alpha\beta})
    \label{g}
\end{equation}
It is clear that when $\lambda=0$, above field equation reduces to Einstein's general field equation.
Now, to understand the cosmological implications of $f(R,T)$ gravity, we assume Friedmann-Lemaitre-Robertson-Walkar (FLRW) metric in spherical coordinate for flat universe,
\begin{equation}
    ds^2= - dt^2 + a(t)^2 \left[dr^2 + r^2 (d\theta^2 + \sin^2\theta d\phi^2)\right]
    \label{h}
\end{equation}
where a(t) is the scale factor, t being the cosmic time.\\ \\
In order to have the inflation, we introduce a homogeneous scalar field $\phi$ minimally coupled to gravity called inflaton. Then the matter Lagrangian will take the form,

\begin{equation}
     L_m = -\frac{1}{2}g^{\alpha\beta}\partial_{\alpha}\phi \partial_{\beta}\phi - V(\phi) = \frac{1}{2}\dot \phi^2 - V(\phi)
    \label{i}
\end{equation}
where $V(\phi)$ is the potential of the scalar field. Then the energy-momentum tensor takes the form,
\begin{equation}
    T_{\alpha\beta} = \partial_{\alpha}\phi \partial_{\beta}\phi + g_{\alpha\beta} [\frac{1}{2}\dot \phi^2 - V(\phi)]
    \label{j}
\end{equation}
Now, computing the components of field Eq. \ref{f} yields,
\begin{equation}
    H^2 = \frac{8\pi G}{3}\left[\frac{\dot\phi^2}{2} (1+ 2\lambda) + V(\phi) (1+ 4\lambda)\right]
    \label{k}
\end{equation}
\begin{equation}
    \frac{\ddot a}{a} = - \frac{8\pi G}{3} \left[\dot\phi^2 (1+ 2\lambda) + V(\phi) (1+ 4\lambda)\right]
    \label{l}
\end{equation}
Eq. \ref{k} is known as the modified Friedmann equation and Eq. \ref{l} is called the modified acceleration equation. From here we can also obtain expression for $\dot H$ as,
\begin{equation}
    \dot H = \frac{\ddot a}{a} - H^2 = - \frac{8\pi G}{2} (p^{eff} + \rho^{eff})= - 4\pi G \dot\phi^2(1+ 2\lambda)
    \label{m}
\end{equation}
Both the above Eq. \ref{l} and Eq. \ref{m} are known as modified Friedmann second equations.
Now, the continuity equation or the modified Klein-Gordon equation in this scenario can be written as,
\begin{equation}
    \ddot \phi (1+ 2\lambda) + 3H\dot \phi (1+ 2\lambda) + \frac{dV}{d\phi} (1+ 4\lambda) = 0
    \label{n}
\end{equation}


\section{Warm Inflation in $f(R,T)$ gravity}
In warm inflationary scenario, the inflaton field interacts with other fields present and during the inflation process converts into radiation during the inflationary period.\cite{r7} So, there is an extra term present in the dynamical equations of the inflaton field due to radiation. The equations that completely specifies the dynamics in the warm inflaton scenario are

\begin{equation}
    H^2=\frac{1}{3M_P^2}(\rho_\phi+\rho_\gamma)
    \label{1}
\end{equation}
\begin{equation}
    \dot{\rho}_\phi+3H(\rho_\phi+P_\phi)=-\Gamma \dot{\phi}^2
    \label{2}
\end{equation}
\begin{equation}
    \dot{\rho}_\gamma+4H\rho_\gamma=\Gamma \dot{\phi}^2
    \label{3}
\end{equation}
Where $\rho_\phi$, $\rho_\gamma$ and $P_\phi$ are the energy density of the scalar field, energy density of the radiation field and pressure of the scalar field. $\Gamma $ is the dissipation coefficient which describe the decay of inflaton into radiation during inflationary phase.

 From equations Eq.~(\ref{d}) and ~(\ref{g}), the energy density and pressure for scalar field are found as
 \begin{equation}
     \rho_\phi=(1+2\lambda)\frac{\dot{\phi}^2}{2}+(1+4\lambda)V(\phi)
     \label{4}
 \end{equation}
\begin{equation}
     P_\phi=(1+2\lambda)\frac{\dot{\phi}^2}{2}-(1+4\lambda)V(\phi)
     \label{5}
 \end{equation}
 Now, equation of motion of inflaton field can be obtained by substituting Eqs.~(\ref{4}) and ~(\ref{5})  into Eq.~(\ref{2}) , 
 \begin{equation}
     (1+2\lambda)\ddot \phi + 3H\dot \phi (1+ 2\lambda+r) + (1+ 4\lambda) \frac{dV}{d\phi}= 0
     \label{6}
 \end{equation}
Here, we restrict our study under the strong dissipative case $r>1$. 
 
 
 \subsection{Slow-Roll parameters}
 Inflation takes place when the potential energy dominates over both the kinetic energy of the inflationary field and the energy density of the radiation field and also it is assumed that radiation production is quasi-stable. This approximation is called slow-roll approximation.\cite{r7} The slow-roll approximations here leads to the conditions
 \begin{equation}
     \frac{(1+2\lambda)}{2}\dot{\phi}^2+\rho_\gamma<<(1+4\lambda)V(\phi)
     \label{7}
\end{equation}
 \begin{equation}
    (1+2\lambda)\ddot \phi << 3H\dot \phi (1+ 2\lambda+r) 
    \label{8}
 \end{equation}
 \begin{equation}
    \dot{\rho}_\gamma<< 4H\rho_\gamma 
    \label{9}
\end{equation}
 In strong dissipative regime warm inflation Eq.~(\ref{1}),~(\ref{6}) and ~(\ref{3}) can be written as
 \begin{equation}
    H^2=(1+4\lambda)\frac{V(\phi)}{3M_P^2}
    \label{10}
\end{equation}
 \begin{equation}
       3H\dot \phi (1+ 2\lambda+r) +(1+4\lambda)\frac{dV}{d\phi}= 0
       \label{11}
 \end{equation}
 \begin{equation}
     \rho_\gamma=\frac{\Gamma \dot{\phi}^2}{4H} =C\tau^4
     \label{12}
 \end{equation}
 Where $C=\frac{g_*\pi^2}{30}$ is the Stefan-Boltzmann constant and $g_*$ is the number of degrees of freedom for the radiation at temperature $\tau$. ( in calculation we will take C = 70 for $ g_*= 200$)
 
 The slow-roll approximation can be parameterized by a set of slow-roll parameters $\epsilon$, $\eta$ and $\beta$ which are defined as
  \begin{equation}
    \epsilon=-\frac{\dot{H}}{H^2} 
    \label{13}
 \end{equation}
 \begin{equation}
  \eta=-\frac{\ddot{H}}{2H\dot{H}}
  \label{14}
  \end{equation}
   \begin{equation}
      \beta=-\frac{\dot{\rho}_\gamma}{H\rho_\gamma}
      \label{15}
  \end{equation} 
  In terms of potential $V(\phi)$ of the scalar field, the slow roll parameters can be expressed as
   \begin{equation}
    \epsilon=\frac{M_P^2}{2(1+2\lambda+r)} \left(\frac{V^\prime}{V}\right)^2
    \label{16}
 \end{equation}
 \begin{equation}
  \eta= \frac{M_P^2}{(1+2\lambda+r)} \left(\frac{V^{\prime\prime}}{V}\right) 
  \label{17}
 \end{equation}
   \begin{equation}
      \beta= \frac{M_P^2}{(1+2\lambda+r)} \left(\frac{\Gamma^\prime V^\prime}{\Gamma V}\right)
      \label{18}
  \end{equation}
  The slow-roll conditions to be satisfied by the slow roll parameters are therefore
  $\epsilon<<1$, $|\eta|<<1$ and $|\beta|<<1$.\\
 
 The number of e-foldings is defined as
 \begin{eqnarray}
N&=&\int^{t_2}_{t_1} H dt=\int^{\phi_f}_{\phi_i}\frac{H}{\dot{\phi}}d\phi \nonumber \\ &=& -\frac{1}{M_P^2}\int^{\phi_f}_{\phi_i}(1+2\lambda+r)\frac{V}{V^\prime} d\phi
\label{19}
\end{eqnarray}


\subsection{Perturbation Calculation}
 We consider the perturbations of the FRW metric in the spatially flat gauge
which can be expressed as
\begin{equation}
      ds^2=-(1+2A)dt^2+2a(t)\partial_iBdtdx^i+a^2(t)\delta_{ij}dx^idx^j
      \label{200}
\end{equation}
 Here $A(x,t)$ and $B(x,t)$ are scalar perturbations of the metric. Under this perturbation, we split, $\phi(x,t)= \phi(t)+\delta\phi(x,t)$, where $\delta\phi(x,t)$ is the linear response due to the thermal stochastic noise. Using the slow-roll conditions, we get the perturbed equation of inflaton field in the spatially flat gauge in the momentum space: \begin{eqnarray}
 &&(1+2\lambda)\ddot{\delta\phi}_k+3H(1+2\lambda+r)\dot{\delta\phi}_k+\frac{k^2}{a^2}(1+2\lambda)\delta\phi_k=(1+2\lambda)\dot{\phi}\dot{A}\nonumber
 \\&+&\frac{k^2}{a^2}(1+2\lambda)\dot{\phi}B-(2(1+4\lambda)V^\prime+\Gamma\dot{\phi})A-\delta\Gamma\dot{\phi}
 \label{201}
 \end{eqnarray}
 And the perturbed Einstein equation becomes
 \begin{equation}
      3H^2A+\frac{k^2}{a}HB=-4\pi G\delta\rho
      \label{202}
  \end{equation}
  \begin{equation}
      HA=4\pi G (\rho+p)\delta u 
      \label{203}
  \end{equation}
  The expressions of $A$ and $B$ from Eqs.~(\ref{202}) and ~(\ref{203}) are substituted into Eq.~(\ref{201}) to eliminate the metric perturbation in the perturbed equation of inflation field. Thus Eq.~(\ref{201}) becomes
\begin{eqnarray}
 &&(1+2\lambda)\ddot{\delta\phi}_k(t)+3H(1+2\lambda+r)\dot{\delta\phi}_k(t)+\frac{k^2}{a^2}(1+2\lambda)\delta\phi_k(t)=\xi_k(t) 
 \label{205}
 \end{eqnarray}
where thermal stochastic noise source $\xi_k(t)$ is introduced to describe the thermal fluctuations. Now, in the slow roll regime, the term $ \ddot{\delta\phi}_k$ can be neglected. Thus the Eq.~(\ref{205}) becomes
\begin{eqnarray}
 && 3H(1+2\lambda+r)\dot{\delta\phi}_k(t)+\frac{k^2}{a^2}(1+2\lambda)\delta\phi_k(t)=\xi_k(t) 
 \label{206}
 \end{eqnarray}
The solution of the Eq.~(\ref{206}) is
  
 \begin{equation}
   \delta\phi_k(t)=\frac{1}{3H(1+2\lambda+r)}\exp{\left[-\frac{t}{\tau}\right]} \int_{t_0}^t\exp{\left[\frac{t^\prime}{\tau}\right]}\xi_k(t^\prime)dt^\prime+ \delta\phi_k(t_0)\exp{\left[-\frac{t-t_0}{\tau}\right]}
 \end{equation}
 Where,
  $ \tau(\phi)=\frac{3H (1+2\lambda+r)}{(1+2\lambda) \frac{k^2}{a^2}}= \frac{3H (1+2\lambda+r)}{(1+2\lambda)k_p^2}$ and $k_p$ is the physical wave number. So, when $k_p$ will increase, the relaxation rate will be faster. If $k_p$ of one mode of $\delta\phi_k$  is smaller than the freeze-out physical wave number $k_F$, then the mode will not thermalize  during a Hubble time.  So, the freeze-out physical wave number $k_F$ is   
 \begin{equation}
     k_F=\sqrt{\frac{3H^2(1+2\lambda+r)}{1+2\lambda} }
     \label{207}
 \end{equation}

Thus, the fluctuations of $\phi$ in warm inflation in f(R,T) gravity
\begin{eqnarray}
 \delta\phi^2&=&\frac{k_F T}{2\pi^2}\nonumber\\&=&\frac{H \tau}{2\pi^2}\sqrt{\frac{3(1+2\lambda+r)}{1+2\lambda}}
 \label{20}
 \end{eqnarray}
 The power spectrum for the scalar fluctuations has the form
 \begin{equation}
     P_R=\left(\frac{H}{\dot{\phi}}\right)^2\delta\phi^2
     \label{21}
 \end{equation}
Using Eqs.~(\ref{10}), ~(\ref{11}) and ~(\ref{20}) in Eq.~(\ref{21}), the expression for $P_R$ in f(R,T) gravity is obtained as
\begin{eqnarray}
P_R &=&\frac{H^3 \tau}{2\pi^2 \dot{\phi}^2}\sqrt{\frac{3(1+2\lambda+r)}{1+2\lambda}}\nonumber\\
&=& \frac{\tau (1+2\lambda+r)^\frac{5}{2}V^\frac{5}{2}(1+4\lambda)^\frac{1}{2}}{2\pi^2 M_P^5 V^{\prime 2}(1+2\lambda)^\frac{1}{2}}
\label{22}
\end{eqnarray}
The above expression for $ P_R$ goes back to the expression based on GR in the limit $\lambda\rightarrow 0$.\\

The power spectrum for the tensor perturbation is

\begin{eqnarray}
P_T&=&\frac{16 H^2}{\pi M_P^2}\nonumber \\
&=& \frac{16 (1+4\lambda)V}{3\pi M_P^4}
\label{23}
\end{eqnarray}
The spectral index $n_s$ and tensor to scalar ratio $R$ can be calculated from Eq. \ref{82} and \ref{83}.


\subsection{Dissipation coefficient}
Dissipation coefficient may be a constant, function of scalar field, the temperature of thermal bath or both temperature and scalar field.\cite{r44} A general expression of dissipation coefficient is
\begin{equation}
    \Gamma(\phi,\tau)=\Gamma_0\frac{\tau^m}{\phi^{m-1}}
    \label{26}
\end{equation}
where $m$ is an integer and $\Gamma_0$ is a dimensionless constant connected to the dissipative microscopic dynamics.\cite{r44} It depends on the couplings and multiplicities of the super fields $X$ and $Y$. The inflaton is coupled with the  bosonic and fermionic components of a super field, $X$, which subsequently decay into the scalar and fermionic components of the super field, $Y$ which again thermalise and give rise to thermal bath. $\Gamma_0$ becomes very large with the increase in the number of decaying fields.\cite{r44}

Different expressions for $\Gamma$ can be obtained for different values of $m$. When $m=-1$, $\Gamma=\Gamma_0\frac{\phi^2}{\tau}$ and this form corresponds to non-SUSY case, when $m=0$,  $\Gamma=\Gamma_0\phi$ that corresponds to SUSY case. When $m=1$,  $\Gamma=\Gamma_0\tau$ represents the high temperature SUSY case and for $m=3$,  $\Gamma=\Gamma_0\frac{\tau^3}{\phi^2}$ corresponds to the low temperature SUSY case.
In this study, we consider two forms of $\Gamma$ 
\begin{itemize}
    \item $\Gamma =\Gamma_*$ = constant
\end{itemize}
\begin{itemize}
    \item $\Gamma=\Gamma_0\frac{\tau^3}{\phi^{2}}$
\end{itemize}

\section{Case study with Chaotic and Natural potentials}

\subsection{Case I: $\Gamma=\Gamma_*$} 

\subsubsection{Chaotic Potential}
In this section we consider chaotic potential\cite{r45} which has the power law form:
\begin{equation}
    V=\Lambda M_P^4\left(\frac{\phi}{M_P}\right)^n
    \label{27}
\end{equation}
where $n$ is the power index and $\Lambda$ is the dimensionless coupling constant. Chaotic potential models are also known as Large Field Inflation (LFI) models. The term chaotic manifests the initial state of the universe. The index parameter $n$ can be a positive number as well as a rational number also. However, the acceptable range is $0.2<n<5$ because models with $n>5$ are ruled out and $n=0$ case is not possible since potential can't be completely flat.\cite{r46} In GR, these models generally give very high tensor-to-scalar ratio as a results they are rejected by the observational data. For example, Planck 2018 data strongly disfavors chaotic models with $n\ge 2$.\cite{r5}

Now, with this potential slow-roll parameters can be expressed as
\begin{equation}
 \epsilon=\frac{M_P^2}{2(1+2\lambda+r)} \left(\frac{n}{\phi}\right)^2   \label{28}
\end{equation}
\begin{equation}
 \eta=\frac{M_P^2}{(1+2\lambda+r)} \frac{n(n-1)}{\phi^2}   
 \label{29}
\end{equation}
 
\begin{figure}
\centerline{\includegraphics[width=8cm]{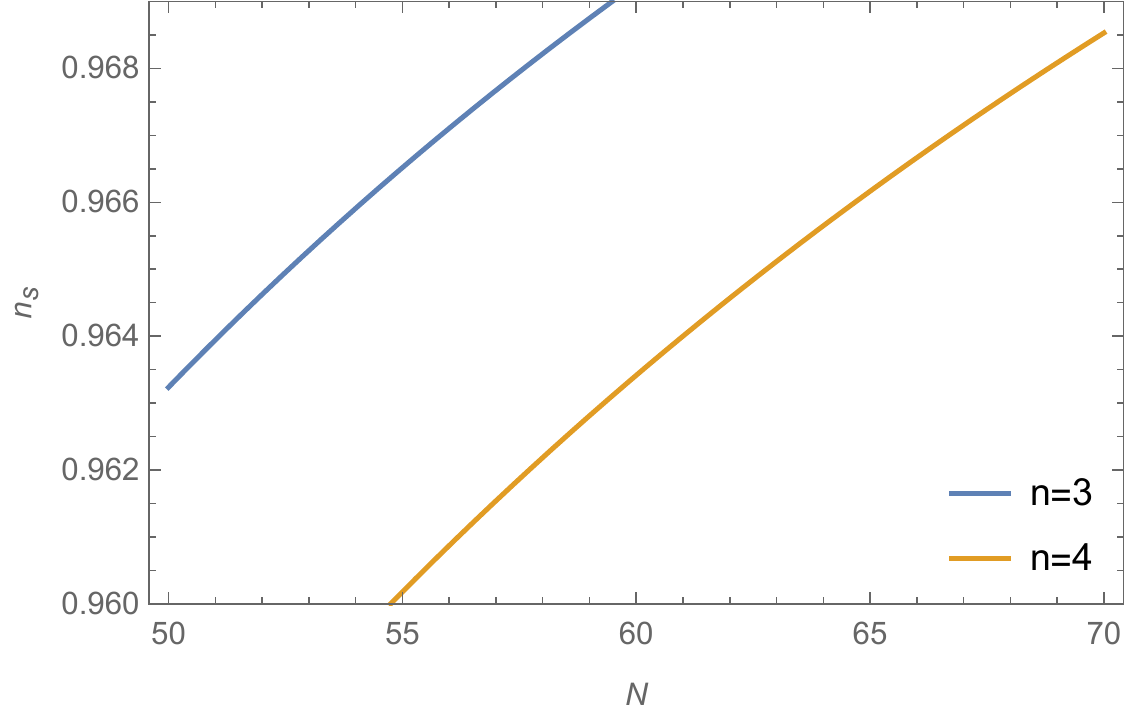}}
\caption{ The spectral index $n_s$ versus $N$ for $n=3$ and $n=4$ \label{f1}}
\end{figure}

Here $\beta=0$ because $\Gamma$ is constant in this case. Warm inflation ends when either the slow-roll conditions are violated i.e. when either of the two parameters $\epsilon$ and $\eta$ becomes of the order of unity earlier. In this case $\eta$ reaches unity earlier than $\epsilon$. So, from the equation $\eta=1$, the final field value $\phi_f$ can be calculated which is
\begin{equation}
 \phi_f=\sqrt{\frac{n(n-1)}{1+2\lambda+r}}M_P  
 \label{30}
\end{equation}
Using the chaotic potential in Eq.~(\ref{27}), we obtain the number of e-folds from Eq.~(\ref{19}) as
\begin{eqnarray}
N=\frac{(1+2\lambda+r)}{2 n M_P^2}(\phi_i^2-\phi_f^2)
\label{31}
\end{eqnarray}
From equation Eq.~(\ref{30}) and ~(\ref{31}) we obtain $\phi_i$ as
\begin{equation}
   \phi_i=\sqrt{\frac{n M_P^2}{1+2\lambda+r}(2N+n-1)}
   \label{71}
\end{equation}
The scalar power spectrum can be calculated using Eq.~(\ref{10}), ~(\ref{11}), ~(\ref{12}) and ~(\ref{27}) and in Eq.~(\ref{22}) as follows
 \begin{equation}
    P_R=\left(\frac{0.035858}{M_P^\frac{3}{2}M_P^\frac{3n}{4}}\right)\left(\frac{\Lambda(1+4\lambda)}{1+2\lambda)}\right)^\frac{1}{2}\left(\frac{1+2\lambda+r}{n}\right)^2\left(\frac{\Lambda n^2(1+4\lambda)r}{C}\right)^\frac{1}{4}\phi_i^{\frac{3}{2}+\frac{3n}{4}}
    \label{32}
\end{equation}
With the help of Eq.~(\ref{71}), scalar power spectrum can be written in terms of $N$
\begin{eqnarray}
P_R&=&\left(\frac{0.035858}{M_P^\frac{3}{2}M_P^\frac{3n}{4}}\right)\left(\frac{\Lambda(1+4\lambda)}{1+2\lambda)}\right)^\frac{1}{2}\left(\frac{1+2\lambda+r}{n}\right)^2\left(\frac{\Lambda n^2(1+4\lambda)r}{C}\right)^\frac{1}{4}\nonumber \\  &&\left(\frac{n M_P^2}{1+2\lambda+r}(2N+n-1) \right) ^{\frac{3}{4}+\frac{3n}{8}}
\label{33}
\end{eqnarray}
From Eq.~(\ref{82}) and ~(\ref{33}), the spectral index in terms of no of e-folds $N$ and potential parameter $n$ can be expressed as
\begin{equation}
    n_s=1-2\left(\frac{3}{4}+\frac{3n}{8}\right)\left(\frac{1}{2N+n-1}\right)
    \label{34}
\end{equation}
It is seen that spectral index $n_s$ does not depend on the $f(R,T)$ model parameter $\lambda$ whereas it depends only on $N$ and $n$. This is similar to the case of standard GR. For $n = 2/3, 1$ and 2, the spectral index does not remain in the Planck 2018 bound. Whereas for $n = 3$ and 4, spectral index remains in the Planck 2018 bound for $N = 50/60/70$ e-folding. The dependence of $n_s$ on the no. of e-folds for $n=3$ and $n=4$ is shown in Fig.~\ref{f1}

\begin{figure}[h!]
 \centering
 \begin{subfigure}[h]{0.49\textwidth}
     \centering
     \includegraphics[width=\textwidth]{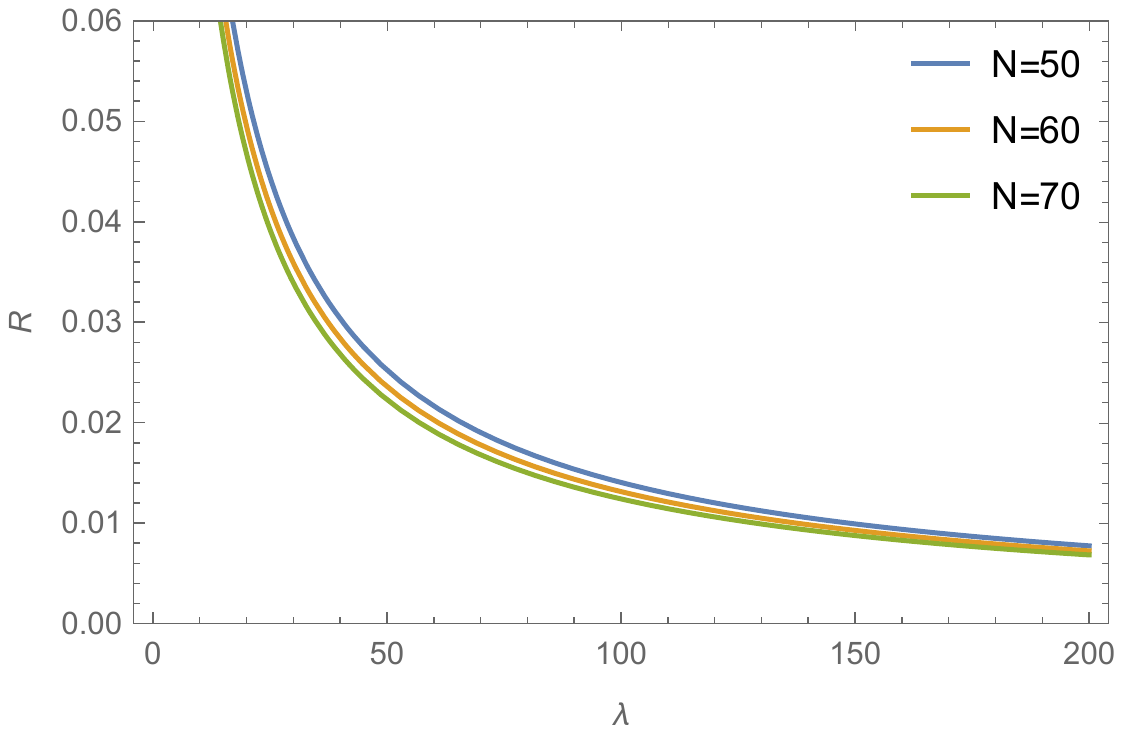}
     \caption{$n=3$}
 \end{subfigure}
 \hfill
 \begin{subfigure}[h]{0.49\textwidth}
     \centering
     \includegraphics[width=\textwidth]{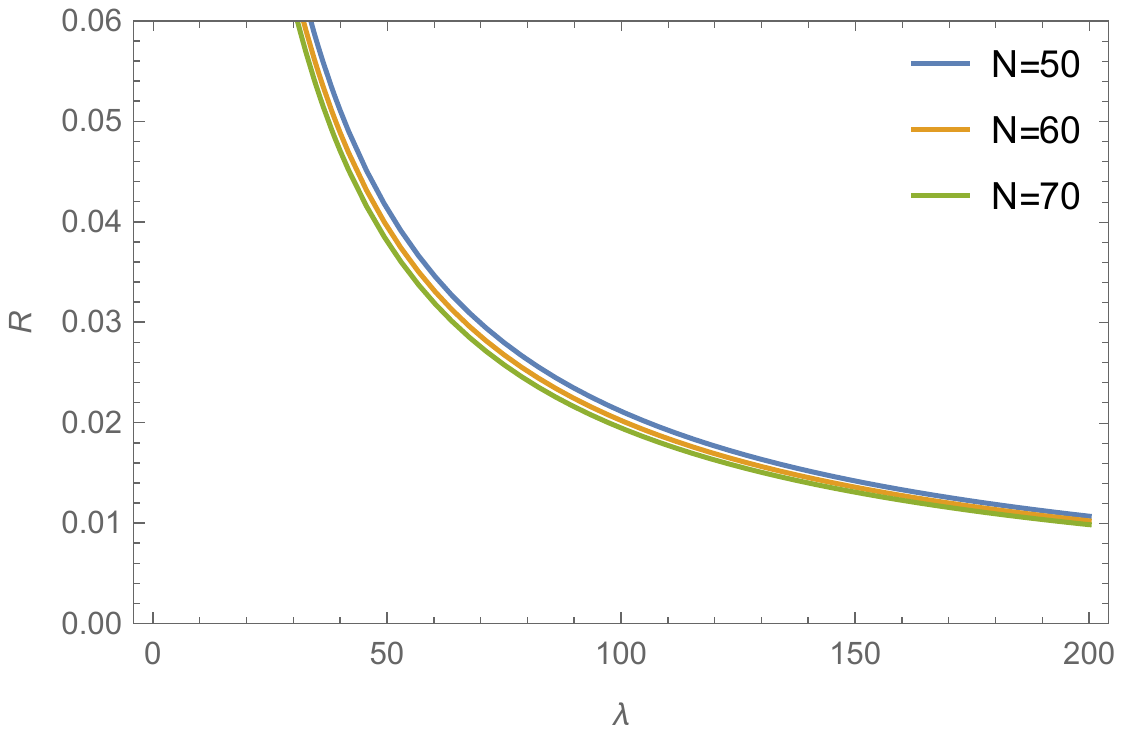}
     \caption{$n=4$}
 \end{subfigure}
 \caption{ The tensor to scalar ratio $R$ versus $\lambda$ for  $\Lambda=0.0000001$, $r=2$, and $C=70$ for different values of the number of e-folds. }
    \label{f2}
 \end{figure}

Tensor power spectrum is obtained from Ens.~(\ref{23}), ~(\ref{27}) and ~(\ref{71}) as
\begin{equation}
    P_T=\frac{16\Lambda(1+4\lambda)}{3\pi M_P^n}\left(\frac{n M_P^2}{1+2\lambda+r}(2N+n-1) \right) ^\frac{n}{2}
    \label{35}
\end{equation}
Using Ens.~(\ref{33}) and ~(\ref{35}) in En.~(\ref{83}), the tensor to scalar ratio in terms of e-folds and model parameters can be expressed as
\begin{eqnarray}
R&=&47.367M_P^{\frac{3}{2}-\frac{n}{4}}\left(\Lambda(1+2\lambda)(1+4\lambda)\right)^\frac{1}{2}\left(\frac{n}{1+2\lambda+r}\right)^2 \left(\frac{C}{\Lambda n^2(1+4\lambda)r}\right)^\frac{1}{4} \nonumber \\ &&\left(\frac{n M_P^2}{1+2\lambda+r}(2N+n-1) \right) ^{\frac{n}{8}-\frac{3}{4}}
\label{36}
\end{eqnarray}
Now, tensor-to-scalar ratio $R$ depends on the model parameter $\lambda$ and other parameters. Considering high dissipation regime with $r=2$ and coupling of the order of $10^{-7}$, we see that $R$ remains in the Planck bound for $n = 2/3, 1, 2, 3$ and 4. However, the cases $n = 2/3, 1, 2$ are ruled out as scalar spectral index for this cases do not match Planck bound. The variation of $R$ with model parameter $\lambda$ for $n=3$ and 4 at different values of e-folds, $N=50, 60$ and $70$ are shown in the Fig.~\ref{f2}. We see that when $n=3$, for $\lambda > 18$ model gives admissible values of $R$ for N = 50/60/70. Whereas when $n=4$ for $\lambda > 30$, model gives admissible values of $R$ for N = 50/60/70.

\subsubsection{Natural Potential}
Natural inflation was originally proposed to solve the fine tuning problem of inflation where the inflaton is interpreted as axion like particle which arises due to a global spontaneous symmetry breaking and the flatness of the potential is protected by the shift symmetry.\cite{r47} The axion is moving on a potential of the form
\begin{equation}
    V(\phi)= \mu^4\left(1+\cos\left(\frac{\phi}{f}\right)\right)
    \label{37}
\end{equation}
Where $f$ is the spontaneous symmetry breaking scale and $\mu$ is the inflationary energy scale with $f>>\mu$. The energy scale $\mu$ depends on the underlying theory and can be range upto GUT scale. In GR, Natural potential is strongly disfavored by Planck 2018 data as it predicts very high tensor-to-scalar ratio for $f \ge 10 M_P$ and low scalar spectral index for $f \sim M_P$.\cite{r5} In f(R,T) gravity also, Natural potential can't meet the observational bounds.\cite{r38} Further, the validity of $f > M_P$ scales in Natural inflation remains questionable. However, it is seen that in warm inflation scenario the symmetry breaking scale can be lower down to GUT scale.\cite{r48} \\
Now the slow-roll parameters are given by
\begin{equation}
    \epsilon=\frac{M_P^2}{2 f^2(1+2\lambda+r)}\left(\frac{1-\cos{\frac{\phi}{f}}}{1+\cos{\frac{\phi}{f}}}\right)
    \label{38}
\end{equation}
\begin{equation}
    \eta=-\frac{M_P^2}{ f^2(1+2\lambda+r)}\left(\frac{\cos{\frac{\phi}{f}}}{1+\cos{\frac{\phi}{f}}}\right)
    \label{39}
\end{equation}
and also $\beta=0$. Inflation ends when the field $\phi$ reaches the value $\phi_f$ by violating one of the conditions $\epsilon<<1$ or $\eta<<1$. It is checked that the second condition violates earlier than the first condition. So, when $\eta=1$, then 
\begin{equation}
    \phi_f=f\left(\pi-\frac{1}{1+\alpha}\right)
    \label{40}
\end{equation}
Where, $\alpha=\frac{M_P^2}{f^2(1+2\lambda+r)}$. Similar to the previous case, using Eqs~(\ref{19}), ~(\ref{37}) and ~(\ref{40}), the value of $\phi_i$ is given by 
\begin{equation}
    \phi_i=2f\sin^{-1}\left[e^{-\frac{\alpha N}{2}}\cos\left({\frac{1}{2+2\alpha}}\right)\right]
    \label{41}
\end{equation}
The expression for scalar power spectrum at $\phi=\phi_i$ can be obtained by using Eq.~(\ref{10}), ~(\ref{11}), ~(\ref{12}) and ~(\ref{37}) in Eq.~(\ref{22}) as
\begin{equation}
    P_R=\left(\frac{.035858 f^\frac{3}{2}\mu^3(1+4\lambda)^\frac{3}{4}r^\frac{1}{4}(1+2\lambda+r)^2}{M_P^\frac{9}{2}C^\frac{1}{4}(1+2\lambda)^\frac{1}{2}}\right)\frac{\left(1+\cos{\left(\frac{\phi_i}{f}\right)}\right)^\frac{3}{2}}{\left(1-\cos{\left(\frac{\phi_i}{f}\right)}\right)^\frac{3}{4}}
    \label{42}
\end{equation}
From Eq.~(\ref{23}) and ~(\ref{37}), tensor power spectrum is given by
\begin{equation}
    P_T=\frac{16\mu^4(1+4\lambda)}{3\pi M_P^4}\left(1+\cos{\left(\frac{\phi_i}{f}\right)}\right)
    \label{43}
\end{equation}
Using Eq.~(\ref{82}), ~(\ref{83}), ~(\ref{41}), ~(\ref{42}) and ~(\ref{43}), we obtain the expressions for tensor to scalar ratio $r$ and spectral index $n_s$ in terms of $f$, $\mu$, $\lambda$, and $N$ which we do not present here as the expressions are lengthy.

\begin{figure}
 \centering
 \begin{subfigure}[h]{0.49\textwidth}
     \centering
     \includegraphics[width=\textwidth]{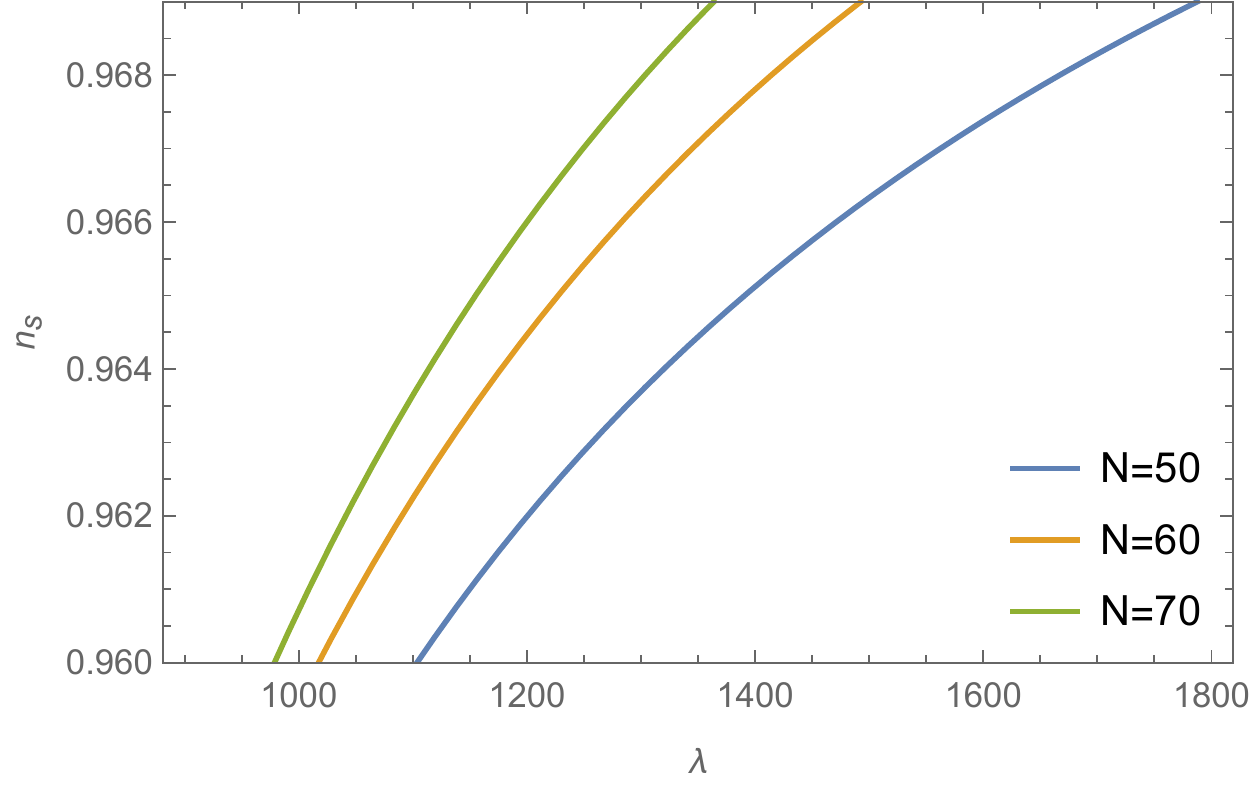}
     \caption{$f=0.1 M_P$}
 \end{subfigure}
 \hfill
 \begin{subfigure}[h]{0.49\textwidth}
     \centering
     \includegraphics[width=\textwidth]{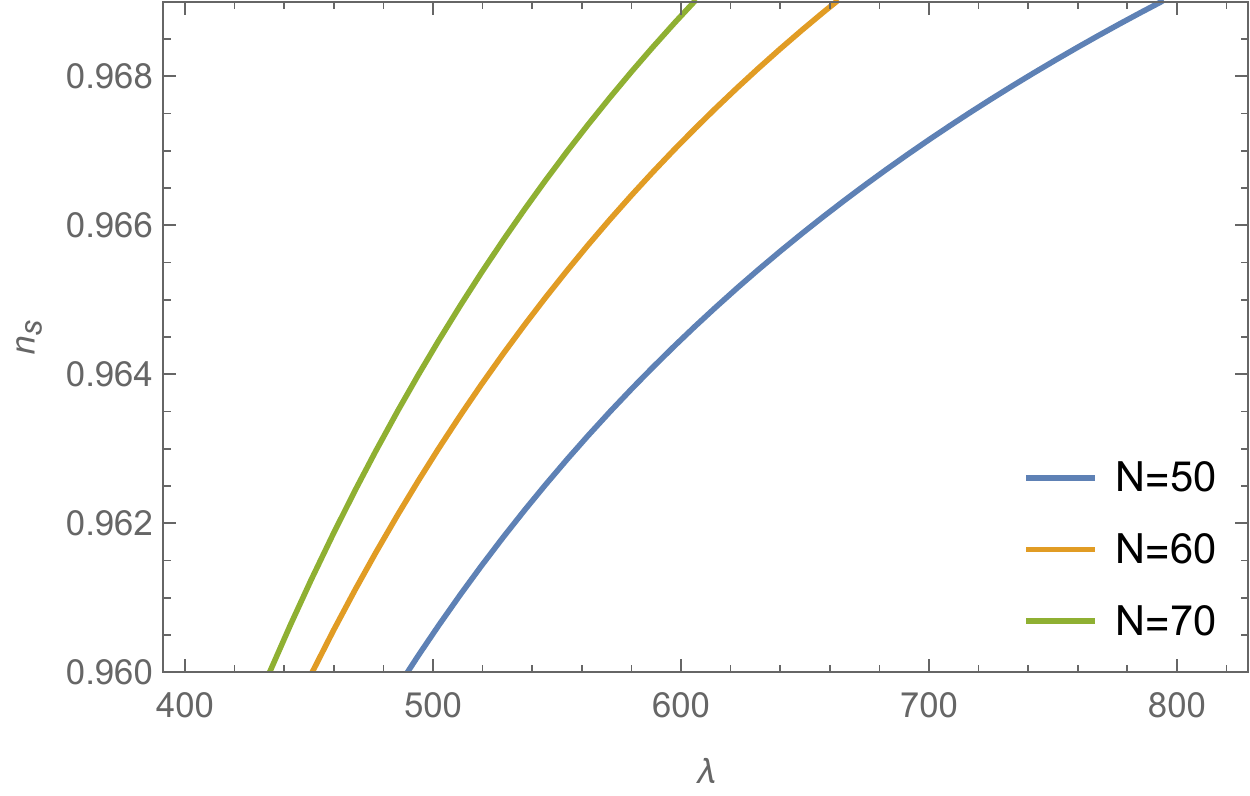}
     \caption{$f=0.15  M_P$}
 \end{subfigure}

 \medskip
 \begin{subfigure}[h]{0.49\textwidth}
     \centering
     \includegraphics[width=\textwidth]{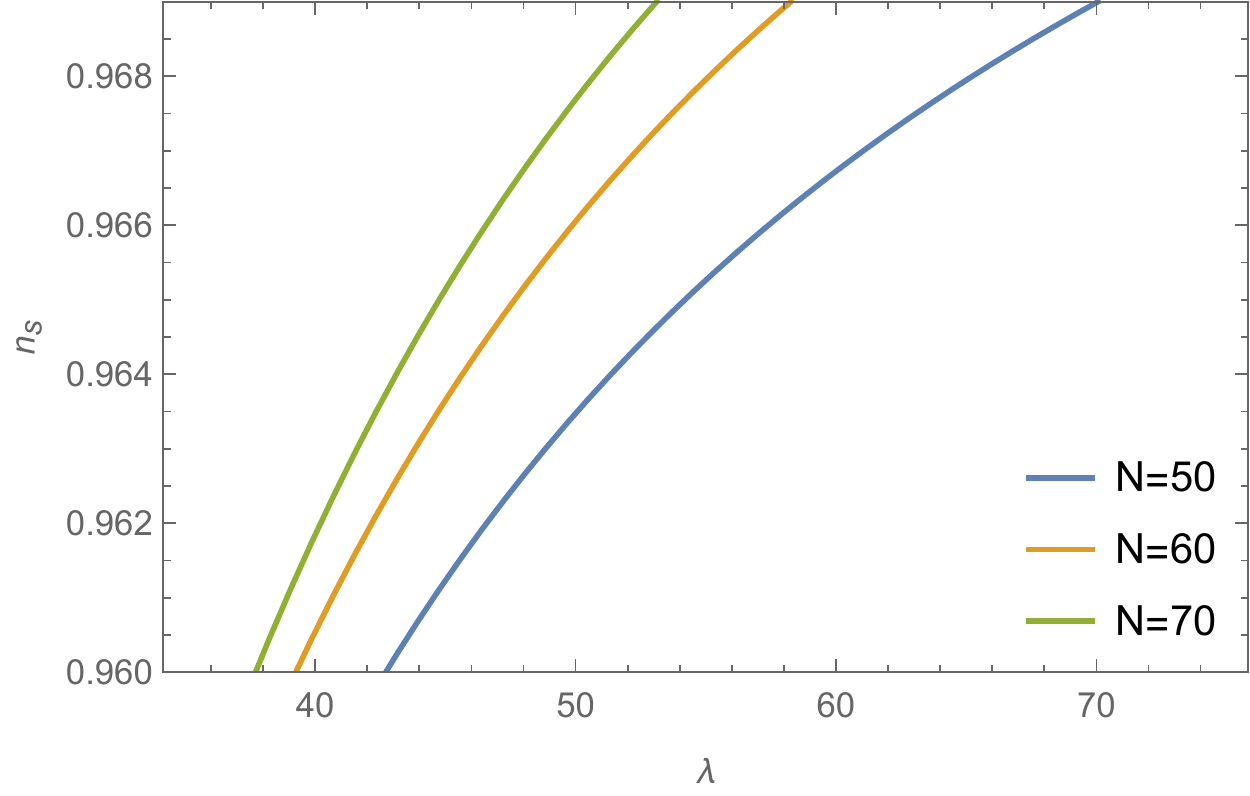}
     \caption{$f=0.5  M_P$}
 \end{subfigure}
 \hfill
 \begin{subfigure}[h]{0.49\textwidth}
     \centering
     \includegraphics[width=\textwidth]{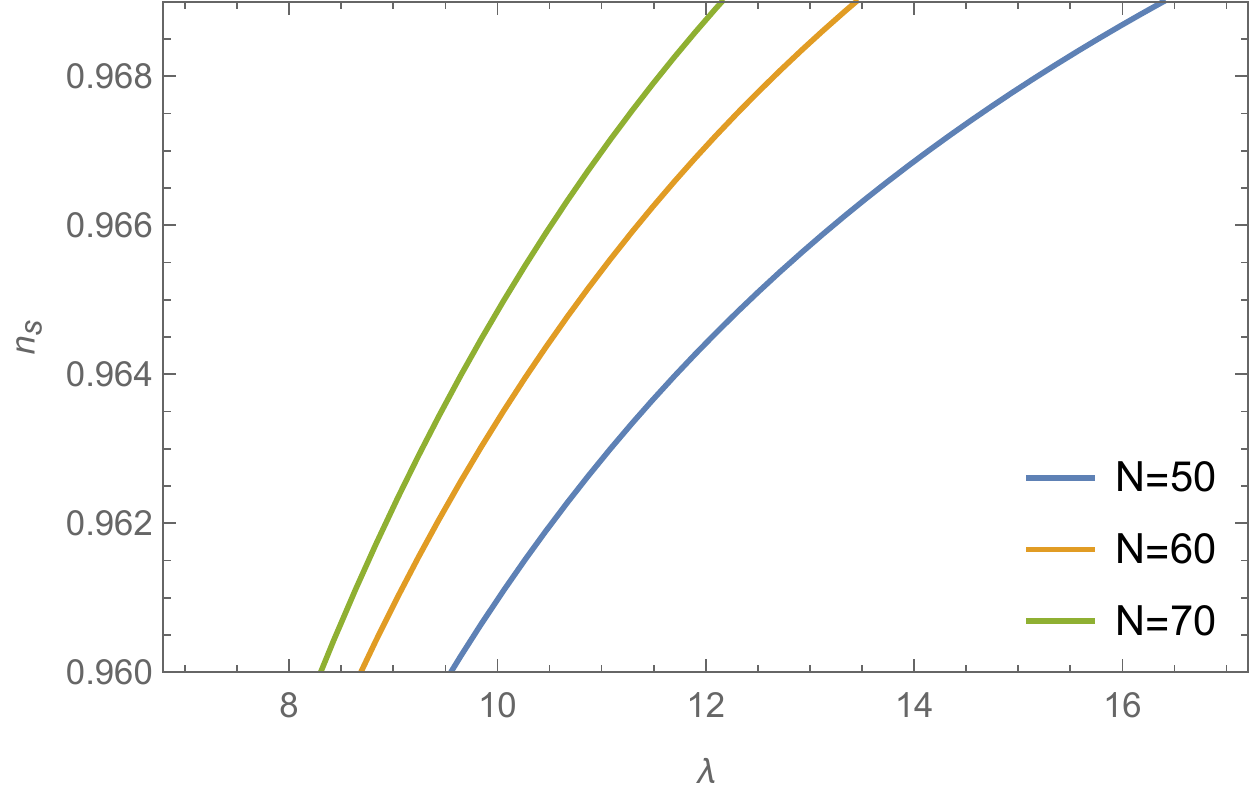}
     \caption{$f=1  M_P$}
 \end{subfigure}
    \caption{ The spectral index $n_s$ versus $\lambda$ for $ \mu=0.01  M_P$, $r=2$ and $C=70$ for different values of the number of e-folds }
    \label{f3}
\end{figure}

Fig.~\ref{f3} shows the dependence of the spectral index on the model parameter $\lambda$ for the different values of the number of e-folds $N$ i.e $N=50,60$ and $70$. It is clear that at $f \sim M_P$ scale, spectral index remains in the Planck 2018 bound. As we go on decreasing $f$ below Planck scale, spectral index remains in the bound provided the model parameter space is shifted to the higher ranges.

\begin{figure}[h!]
 \centering
 \begin{subfigure}[h]{0.49\textwidth}
     \centering
     \includegraphics[width=\textwidth]{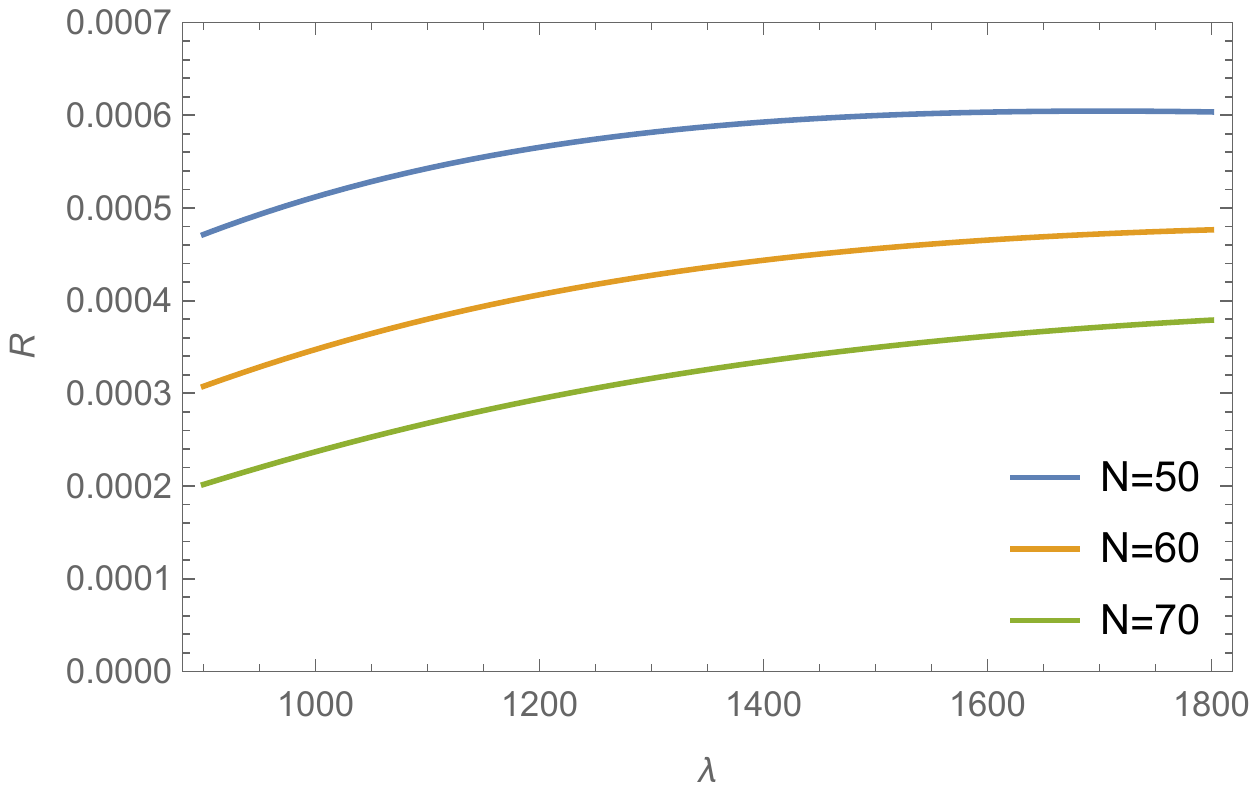}
     \caption{$f=0.1 M_P$}
 \end{subfigure}
 \hfill
 \begin{subfigure}[h]{0.49\textwidth}
     \centering
     \includegraphics[width=\textwidth]{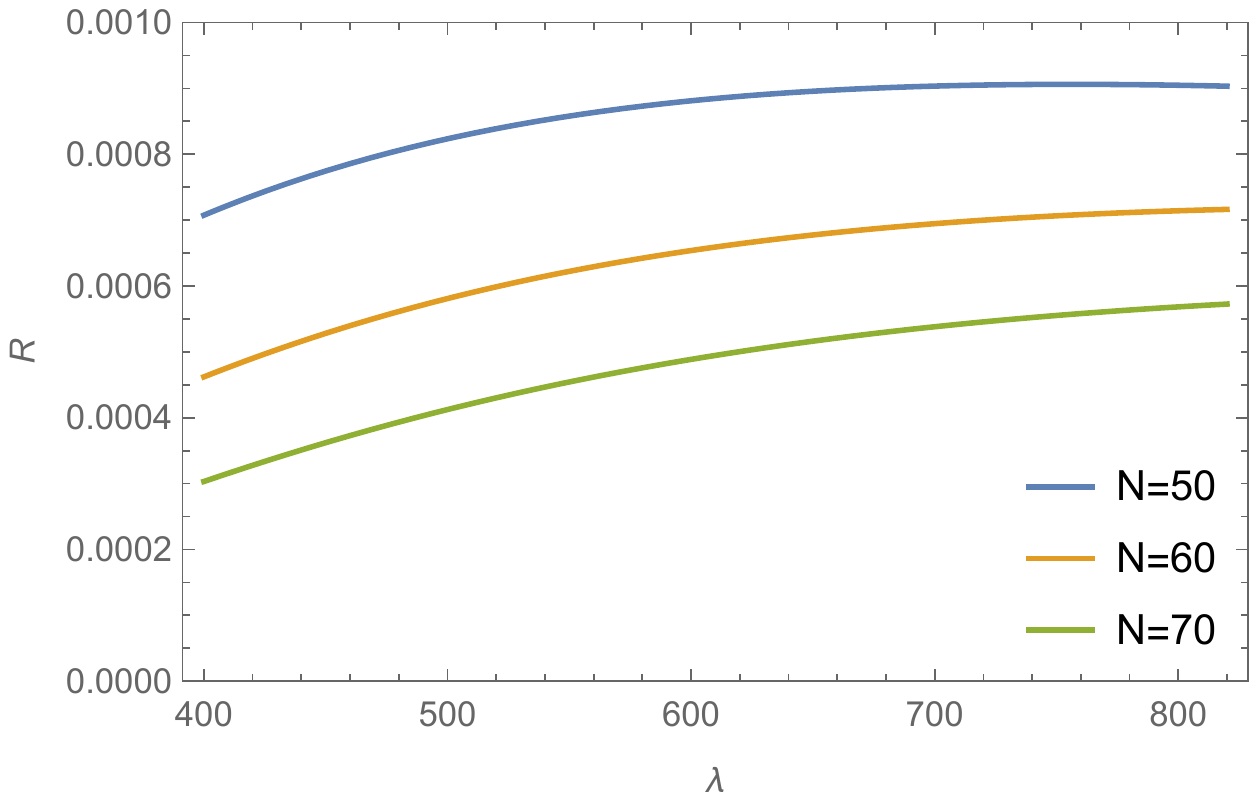}
     \caption{$f=0.15  M_P$}
 \end{subfigure}

 \medskip
 \begin{subfigure}[h]{0.49\textwidth}
     \centering
     \includegraphics[width=\textwidth]{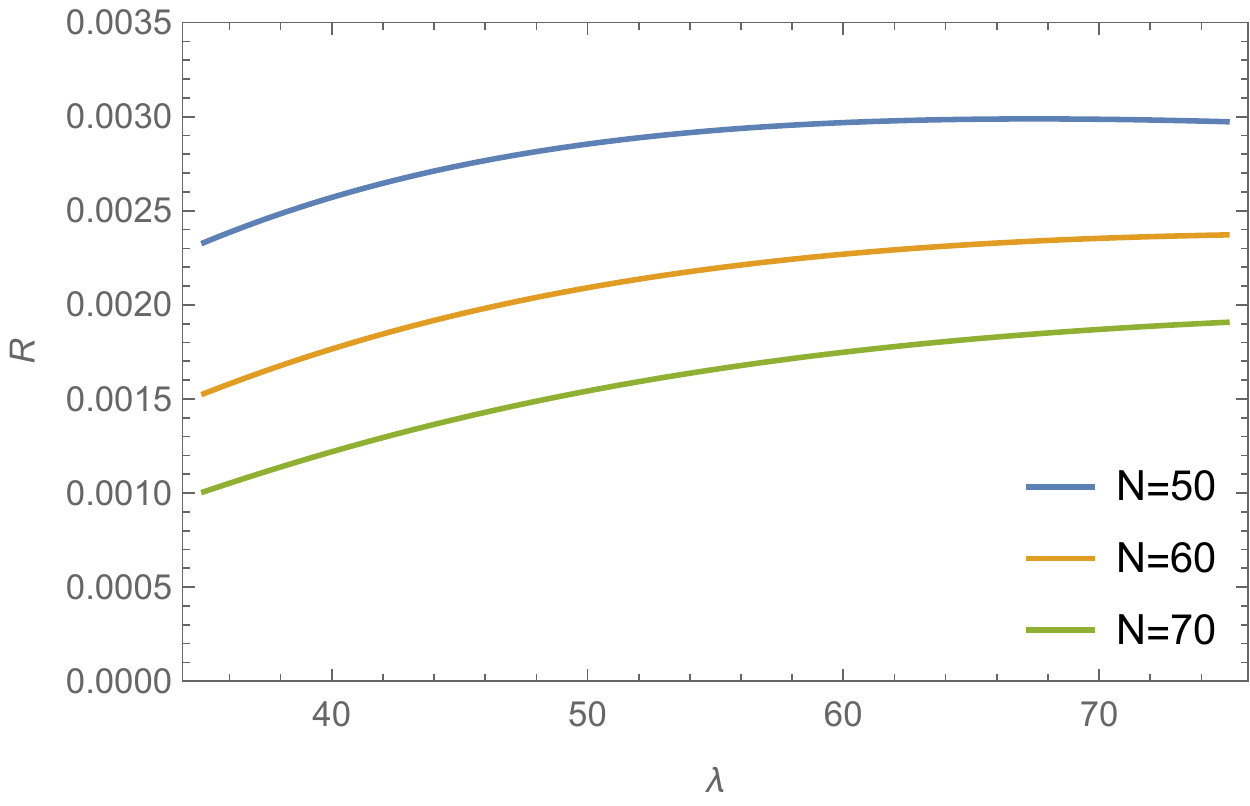}
     \caption{$f=0.5  M_P$}
 \end{subfigure}
 \hfill 
 \begin{subfigure}[h]{0.49\textwidth}
     \centering
     \includegraphics[width=\textwidth]{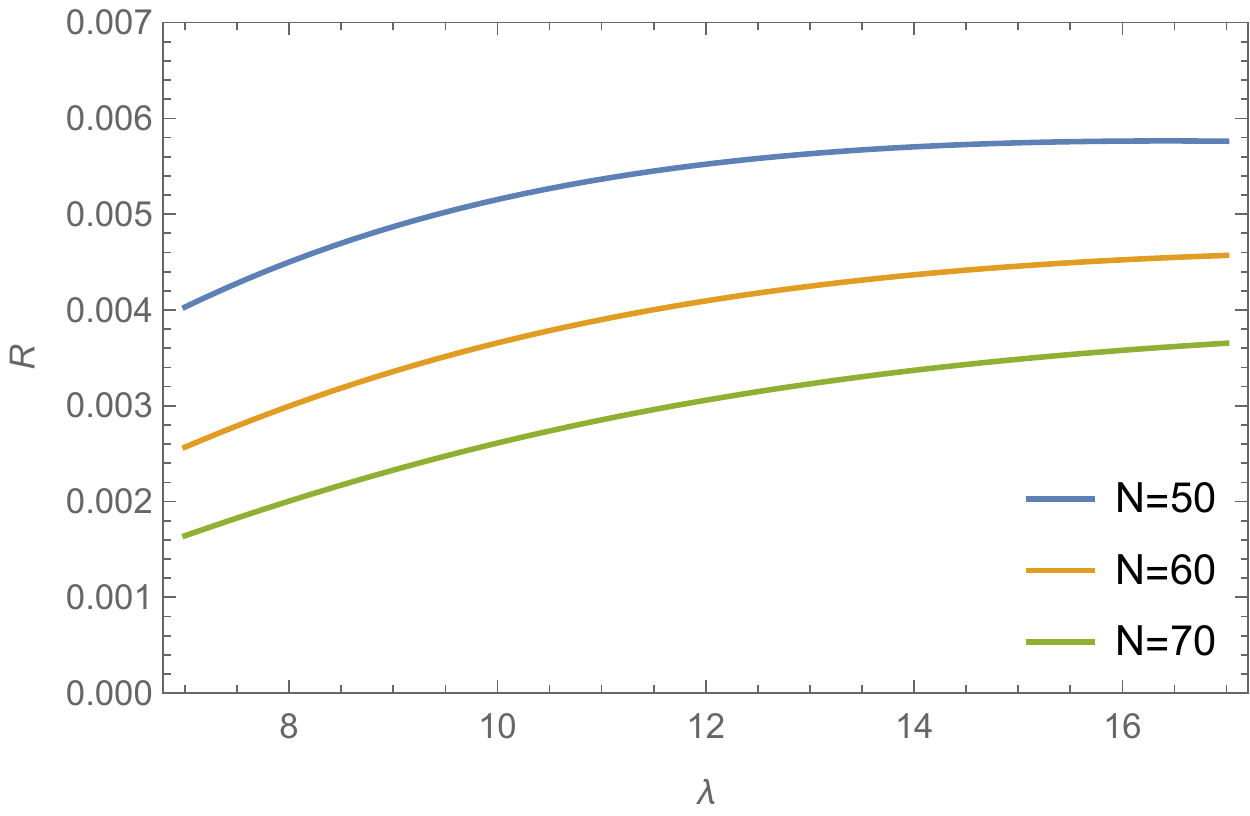}
     \caption{$f=1  M_P$}
 \end{subfigure}
    \caption{ The tensor to scalar ratio $R$ versus $\lambda$ for $\mu=0.01 M_P$, $r=2$ and $C=70$ for different values of the number of e-folds }
    \label{f4}
\end{figure}

  Fig.~\ref{f4} shows the graphical behavior of tensor to scalar ratio $R$ versus model parameter $\lambda$ for the different values of the number of e-folds $N$ i.e $N=50,60$ and $70$.
We see that tensor-to-scalar ratio is of the order of $10^{-3}$ at $f \sim M_P$ scale which is within the Planck 2018 range. Further with the decrease in $f$ scale, $R$ still remains within Planck 2018 bound but the range of the model parameter $\lambda$ increases as we have seen in the case of spectral index.

\subsection{Case II: $\Gamma=\Gamma_0
\frac{\tau^3}{\phi^2}$}
\subsubsection{Chaotic Potential}
With chaotic potential $V=\Lambda M_P^4\left(\frac{\phi}{M_P}\right)^n$, the Friedmann equation can be written as
\begin{equation}
    H^2=(1+4\lambda)\frac{\Lambda}{3}\frac{M_P^2}{M_P^n}\phi^n
    \label{44}
\end{equation}
Using Eq.~(\ref{44}) and the potential, we can express $\dot{\phi}$ as
\begin{equation}
    \dot{\phi}=-\left(\frac{n\Lambda^\frac{1}{2}}{\sqrt{3}}\right)\left(\frac{(1+4\lambda)^\frac{1}{2}}{(1+2\lambda+r)}\right)\left(\frac{M_P^3}{M_P^\frac{n}{2}}\right)\phi^{\frac{n}{2}-1}
    \label{45}
\end{equation}
Combining Eq.~(\ref{44}) and ~(\ref{45}), we get
\begin{equation}
    \frac{H^2}{2\pi\dot{\phi}}=-\left(\frac{\Lambda^\frac{1}{2}}{2\pi\sqrt{3} n}\right)(1+4\lambda)^\frac{1}{2}(1+2\lambda+r)\frac{\phi}{M_P}^{\frac{n}{2}+1}
    \label{46}
\end{equation}
Now, from Eq.~(\ref{12}),
\begin{equation}
    \tau=\left(\frac{3r\dot{\phi}^2}{4C}\right)^\frac{1}{4}
    \label{47}
\end{equation}
Substituting Eq.~(\ref{45}) in Eq.~(\ref{47}), the temperature of the thermal bath is obtained as
\begin{equation}
    \tau=\left(\left(\frac{n^2\Lambda}{4C}\right)\left(\frac{r(1+4\lambda)}{(1+2\lambda+r)^2}\right)\left(\frac{M_P^6}{M_P^n}\right)\phi^{n-2}\right)^\frac{1}{4}
    \label{48}
\end{equation}
From Eq.~(\ref{44}) and ~(\ref{48}), we get
\begin{equation}
    \frac{\tau}{H}=1.2247\left(\frac{r}{C \Lambda(1+4\lambda)}\right)^\frac{1}{4}\left(\frac{n}{1+2\lambda+r}\right)^\frac{1}{2}\left(\frac{\phi}{M_P}\right)^{-\left(\frac{n}{4}+\frac{1}{2}\right)}
    \label{49}
\end{equation}
With the considered form of $\Gamma$ and recalling that $\Gamma=3r H$,
\begin{equation}
 \tau=\left(\frac{3r\phi^2 H}{\Gamma_0}\right)^\frac{1}{3} 
 \label{51}
\end{equation}
Equating Eq. ~(\ref{48}) and ~(\ref{51}),
\begin{equation}
    \frac{\phi}{M_P}=\left(\frac{\Gamma_0^4 n^6 \Lambda (1+4\lambda)}{9 (4C)^3 r (1+2\lambda+r)^6}\right)^\frac{1}{14-n}
    \label{52}
\end{equation}
Differentiating Eq.~(\ref{52}) with respect to $N$,
\begin{equation}
    \frac{ d r}{d N}=\left(\frac{n r(14-n)}{(1+2\lambda+7r)}\right)\left(\frac{9(4C)^3 r (1+2\lambda+r)^6}{\Gamma_0^4 n^6 \Lambda (1+4\lambda)}\right)^\frac{2}{14-n}
    \label{53}
\end{equation}
Using
\begin{equation}
    \frac{d \phi}{d N}=\frac{\dot{\phi}}{H}=-\frac{n M_P^2}{(1+2\lambda+r)\phi}
\end{equation}\\
Now the expressions for scalar power spectrum and tensor power spectrum in terms of $\lambda$, $\Lambda$, $\Gamma_0$, $C$, $n$ and $r$ are
\begin{equation}
  P_R=\frac{(1+2\lambda+r)^2(1+4\lambda)^\frac{3}{4}r^\frac{1}{4}\Lambda^\frac{3}{4}}{2\sqrt{2}\pi^2 C^\frac{1}{4} n^\frac{3}{2}(1+2\lambda)^\frac{1}{2}}\left(\frac{\Gamma_0^4 n^6 \Lambda (1+4\lambda)}{9 (4C)^3 r (1+2\lambda+r)^6}\right)^\frac{3n+6}{4(14-n)}  
  \label{54}
\end{equation}
\begin{equation}
    P_T=\frac{16\Lambda(1+4\lambda)}{3\pi}\left(\frac{\Gamma_0^4 n^6 \Lambda (1+4\lambda)}{9 (4C)^3 r (1+2\lambda+r)^6}\right)^\frac{n}{14-n}
    \label{55}
\end{equation}
From Eq.~(\ref{54}) and ~(\ref{55}), spectral index and tensor to scalar ratio are obtained as
\begin{equation}
    n_s=1-\frac{n((4-2 n) (1+2\lambda+r)+(38-13 n) r)}{2 (1+2\lambda+r) (1+2\lambda+7r)} \left(\frac{\Lambda n^6 (1+4\lambda) \Gamma_0^4}{9 (4 C)^3 r (1+2\lambda+r)^6}\right)^{-\frac{2}{14-b}}
    \label{56}
\end{equation}
\begin{equation}
    R=\frac{32\sqrt{2}\pi C^\frac{1}{4} n^\frac{3}{2}\Lambda^\frac{1}{4}(1+2\lambda)^\frac{1}{2}(1+4\lambda)^\frac{1}{4}}{3  r^\frac{1}{4} (1+2\lambda+r)^2}\left(\frac{\Gamma_0^4 n^6 \Lambda (1+4\lambda)}{9 (4C)^3 r (1+2\lambda+r)^6}\right)^\frac{n-6}{4(14-n)}
    \label{57}
\end{equation}
Total number of e-foldings can be obtained by integrating $\frac{d N}{d r}$ from $r_i$ to $r_f$ as
\begin{eqnarray}
    N&=&\left(\frac{\Gamma_0^4 n^6 \Lambda (1+4\lambda)}{9 (4C)^3}\right)^\frac{2}{14-n}\left(\frac{1}{n(14-n)}\right)\left[\left(\frac{14-n}{2(12-n)}\right)\left(\frac{1}{r}\right)^\frac{2}{14-n}\nonumber \right.\\&& 
    \left(\frac{1}{1+2\lambda}\right)^\frac{12}{14-n}\left((n-12)(1+2\lambda+r)\left(\frac{1+2\lambda}{1+2\lambda+r} \right)^\frac{12}{14-n}+(14-n)r\nonumber \right. \\ 
    &-&\left.\left.\frac{6(12-n)r^2}{(13-n)(1+2\lambda)}+\frac{6(26-n)(12-n)r^3}{(14-n)(40-3n)(1+2\lambda)^2} \right)\right]\bigg|^{r_f}_{r_i}
    \label{58}
\end{eqnarray}
$r_f$ corresponds to $\eta=1$, $\eta$ being the largest of the slow-roll parameters.
\begin{equation}
  \frac{M_P^2}{(1+2\lambda+r_f)} \frac{n(n-1)}{\phi^2}=1 
  \label{59}
\end{equation}
Substituting Eq.~(\ref{52}) in Eq.~(\ref{59}), we obtain
\begin{equation}
    r_f(1+2\lambda+r_f)^{\frac{n}{2}-1}=\frac{\Gamma_0^4 n^6 \Lambda (1+4\lambda)}{9(4C)^3(n(n-1))^{7-\frac{n}{2}}}
    \label{60}
\end{equation}
Now, for $n=2$, we have
\begin{equation}
    r_f=\frac{\Gamma_0^4 \Lambda (1+4\lambda)}{9(4C)^3}
    \label{61}
\end{equation}
Substituting the expression of $r_f$ in Eq.~(\ref{58}) and solving this equation numerically, value of $r_i$ is obtained for $N=60$. The parameter space is then fixed so that the spectral index and tensor to scalar ration lie within the PLANCK bound.

The possible values of the spectral index and tensor to scalar ratio at $r=r_i$ for different values of model parameters are presented in Table \ref{ta1} .

\begin{table}[h!]
\centering
{\begin{tabular}{@{}cccccc@{}} \toprule
 $\Lambda$ & $\Gamma_0$ & $\lambda$  & $ r_i$   & $ n_s$  & $ R $ \\
 \midrule
&     & 145 &  5 &   0.9661  &  0.057\\
&     &150  &  5.4 &   0.9648  & 0.055\\
$10^{-6}$ &  5189  & 155 & 5.8  & 0.9630  & 0.053\\
&           & 160 &  6.2 &   0.9616  &  0.051\\
&           & 165 &  6.7 &   0.9602  &  0.049\\
\midrule
&     & 152 &  4.8 &   0.9694  &  0.031\\
&     & 160 &  5.4 &   0.9674  & 0.029\\
$10^{-7}$ & 9227 & 170 & 6  & 0.9649  & 0.027\\
&     & 180 &  6.8 &   0.9630  &  0.025\\
&     & 191 &  7.7 &   0.9604  &  0.024\\
\midrule
&     & 120 &  3.8 &   0.9693  &  0.022\\
&     & 125 &  4.2 &   0.9673  & 0.021\\
$10^{-8}$ & 16410    & 130 & 4.6  & 0.9655  & 0.02\\
&     & 140 &  5.5 &   0.9615  &  0.018\\
&     & 145 &  5.9 &   0.9600  &  0.017\\
\bottomrule
\end{tabular}
\caption{Possible values of the observable for $ n=2$, $C=70$ and $N=60$}
\label{ta1}}
\end{table}

 It is clear from Table \ref{ta1} that $n=2$ model produces results consistent with Planck 2018 data. The range of model parameter is found to be $145 < \lambda < 165$ for $\Lambda = 10^{-6}$, $152 < \lambda < 191$ for $\Lambda = 10^{-7}$ and $120 < \lambda < 145$ for $\Lambda = 10^{-8}$. Further, it is seen that with the increase in model parameter $\lambda$, tensor-to-scalar ratio decreases and with the decrease in the value of coupling constant $\Lambda$, tensor-to-scalar ratio decreases. It is pertinent to mention here that the cases $n = 1, 2/3, 3$ and 4 are either not solvable or does not yield desired results.

\subsubsection{Natural Potential}
 
  With $ V(\phi)= \mu^4\left(1+\cos\left(\frac{\phi}{f}\right)\right)$, the three slow-roll parameters are
\begin{equation}
    \epsilon=\frac{M_P^2}{2 f^2(1+2\lambda+r)}\left(\frac{1-\cos{\frac{\phi}{f}}}{1+\cos{\frac{\phi}{f}}}\right)
    \label{62}
\end{equation}
\begin{equation}
    \eta=-\frac{M_P^2}{ f^2(1+2\lambda+r)}\left(\frac{\cos{\frac{\phi}{f}}}{1+\cos{\frac{\phi}{f}}}\right)
    \label{63}
\end{equation}
\begin{equation}
    \beta=\frac{2 M_P^2}{ f(1+2\lambda+r)\phi}\left(\frac{\sin{\frac{\phi}{f}}}{1+\cos{\frac{\phi}{f}}}\right)
    \label{64}
\end{equation}
The slow-roll parameter that violates the slow-roll condition first gives the final field value $\phi_f$. Initial field value $\phi_i$ can be obtained by taking the number of e-folds $N=60$ and using the equation
\begin{eqnarray}
N&=& \frac{1}{M_P^2}\int^{\phi_i}_{\phi_f}(1+2\lambda+r)\frac{V}{V^\prime} d\phi  \nonumber \\
&=&\frac{f}{M_P^2}\int^{\phi_f}_{\phi_i}(1+2\lambda+r)\frac{1+\cos{\left(\frac{\phi}{f}\right)}}{\sin{\left(\frac{\phi}{f}\right)}} d\phi 
\label{65}
\end{eqnarray}
Similar to the previous case, we get the expression for scalar power spectrum and tensor power spectrum are obtained as
\begin{equation}
      P_R=\left(\frac{.035858 f^\frac{3}{2}\mu^3(1+4\lambda)^\frac{3}{4}r^\frac{1}{4}(1+2\lambda+r)^2}{M_P^\frac{9}{2}C^\frac{1}{4}(1+2\lambda)^\frac{1}{2}}\right)\frac{\left(1+\cos{\left(\frac{\phi}{f}\right)}\right)^\frac{3}{2}}{\left(1-\cos{\left(\frac{\phi}{f}\right)}\right)^\frac{3}{4}}
      \label{66}
\end{equation}
 
\begin{equation}
    P_T=\frac{16\mu^4(1+4\lambda)}{3\pi M_P^4}\left(1+\cos{\left(\frac{\phi}{f}\right)}\right)
    \label{67}
\end{equation}
Now, using Eq.~(\ref{11}), $\dot{\phi}$ can be expressed as
\begin{equation}
    \dot{\phi}=\frac{M_P \mu^2(1+4\lambda)^\frac{1}{2}}{\sqrt{3}(1+2\lambda+r)f}\frac{\sin{\frac{\phi}{f}}}{\left(1+\cos{\frac{\phi}{f}}\right)^\frac{1}{2}} 
    \label{68}
\end{equation}
 Substituting the expression of $\dot{\phi}$ in Eq.~(\ref{12}), the thermal bath temperature turns out as
\begin{equation}
    \tau=\left(\left(\frac{M_P^2 \mu^4}{4C f^2}\right)\left(\frac{r(1+4\lambda)}{(1+2\lambda+r)^2}\right)\left(\frac{\sin^2{\frac{\phi}{f}}}{1+\cos{\frac{\phi}{f}}}\right) \right)^\frac{1}{4}
    \label{69}
\end{equation}
In this case, $\Gamma=\Gamma_0
\frac{\tau^3}{\phi^2}$ giving $\tau=\left(\frac{3r\phi^2 H}{\Gamma_0}\right)^\frac{1}{3} $. Equating this with Eq.~(\ref{69})
\begin{equation}
    r(1+2\lambda+r)^6=\left(\frac{M_P^{10} \mu^4 \Gamma_0^4(1+4\lambda)}{9(4C)^3 f^6}\right)\frac{\left(1-\cos{\frac{\phi}{f}}\right)^3}{\phi^8\left(1+\cos{\frac{\phi}{f}}\right)^2}
    \label{70}
\end{equation}
Solving Eq.~(\ref{70}) and using the solution in Eq.~(\ref{62}),~(\ref{63}) and ~(\ref{64}) the slow-roll parameters are obtained as a function of $ f $, $\lambda$, $\phi$, $C$, $\mu$ and $\Gamma_0$ where $ f $,  $\phi$, and $\mu$ are written in $M_P$ unit.

\begin{figure}[h!]
     \centering
     \begin{subfigure}[b]{0.45\textwidth}
         \centering
         \includegraphics[width=\textwidth]{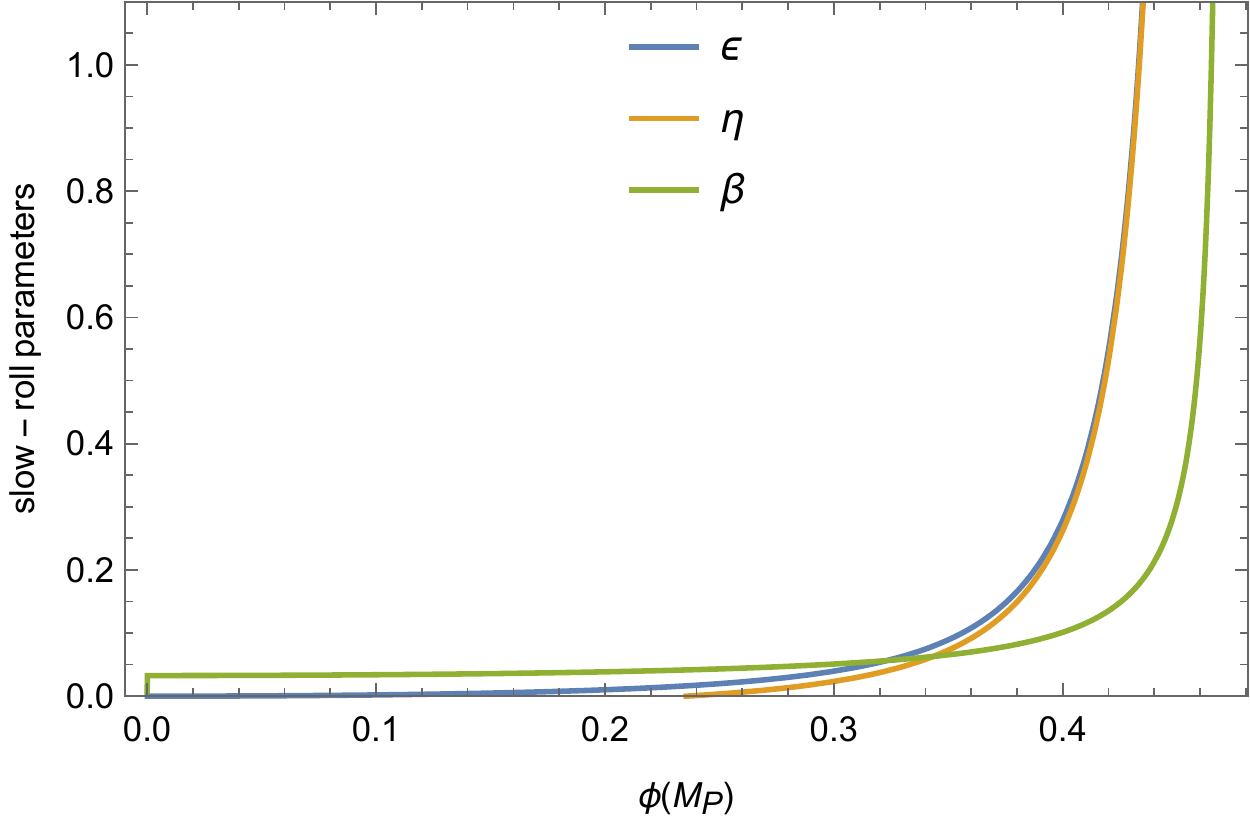}
         \caption{$\lambda=680, \Gamma_0=5\times10^3$}
         \label{fig:y equals x}
     \end{subfigure}
     \hfill
     \begin{subfigure}[b]{0.45\textwidth}
         \centering
         \includegraphics[width=\textwidth]{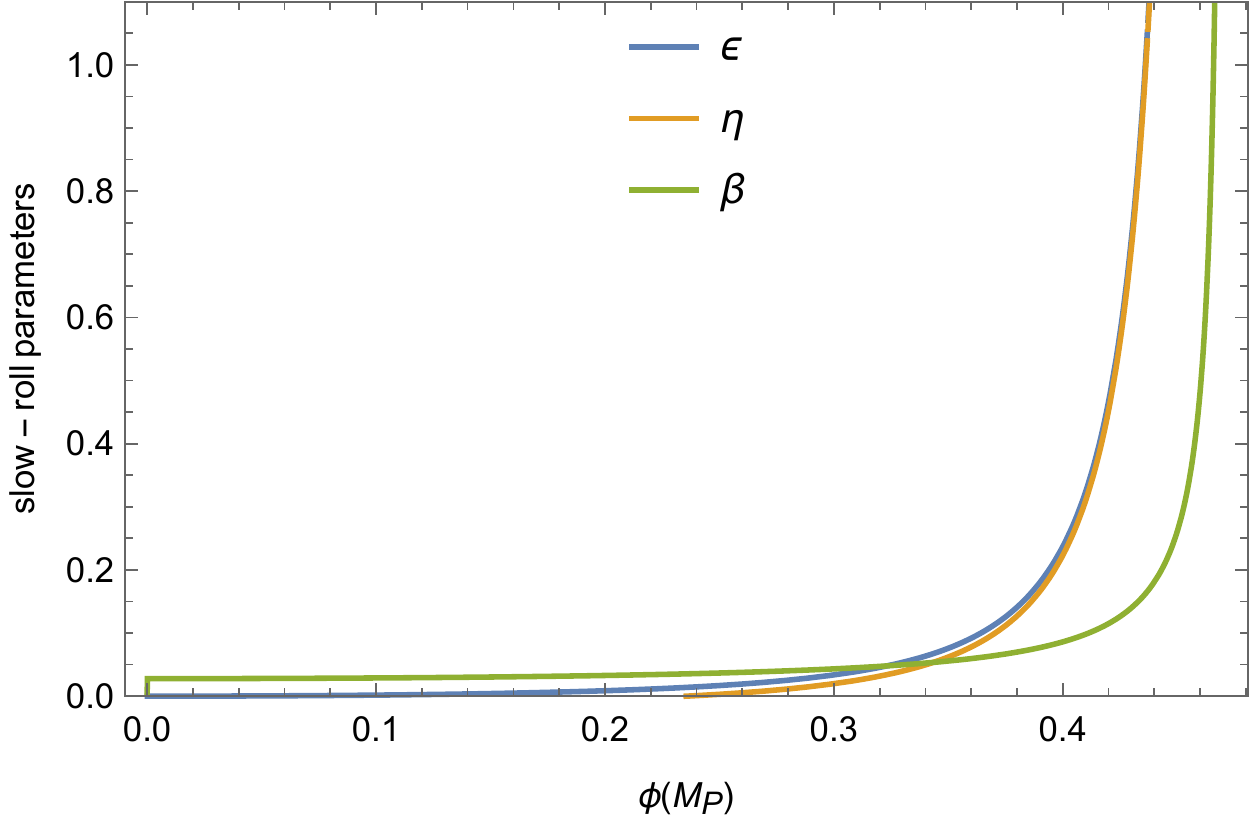}
         \caption{$\lambda=800, \Gamma_0=10^4$}
         \label{fig:three sin x}
     \end{subfigure}
     \caption{Slow-roll parameters versus $\phi$ for $f=0.15 M_P$, $\mu=0.01 M_P$, and $C=70$ for different values of $\lambda$ and $\Gamma_0$ }
        \label{f5}
\end{figure}

 In Fig.~\ref{f5} the variation of  slow-roll parameters with the field value $\phi$ for different values of $\lambda$ and $\Gamma_0$ are shown.

It is found that, in every cases either $\epsilon$ or $\eta $ violates the slow-roll condition first. Either $\eta(\phi_f)=1 $ or $\epsilon(\phi_f)=1 $ as the case may be used to determine the final field value.

The spectral index and tensor to scalar ratio is determined for field values at 60 e-folds before the end of inflation using Eq.~(\ref{82}), ~(\ref{83}),~(\ref{66}) and ~(\ref{67}). The possible values of the spectral index and tensor to scalar ratio for different values of model parameters are presented below.

 \begin{table}[h!]
\centering
{\begin{tabular}{@{}cccccc@{}} \toprule
 $\Gamma_0$ & $\lambda$   & $ n_s$  & $ R $ \\
 \midrule
& 680 &  0.9605   &  0.041\\
$5\times10^3$  &700 &  0.9612   & 0.044\\
&750 &  0.9626   & 0.051\\
&780 &  0.9635  &  0.055\\
\midrule
&   680 &  0.9605   &  0.02\\
$10^4$  &800 &  0.9640   & 0.03\\
&1000 &  0.9676   & 0.04\\
&1150 &  0.9694  &  0.05\\
\midrule
&680 &  0.9604   &  0.002\\
$10^5$ &800 &  0.9640   & 0.003\\
&1000 &  0.9676   & 0.004\\
&1150 &  0.9694  &  0.005\\
\bottomrule
\end{tabular}
\caption{Possible values of the observables for $ f=0.15 M_P$, $\mu=0.01 M_P$, $C=70$ and $N=60$}
\label{ta2}}
\end{table}

From Table 2, it is seen that natural potential can give consistent result at sub planckian scale. The range of the model parameter is found to be $680 < \lambda < 780$ for $\Gamma_0 = 5 \times 10^{3}$, $680 < \lambda < 1150$ for $\Gamma_0 = 10^{4}$ and $680 < \lambda < 1150$ for $\Gamma_0 = 10^{5}$. With the increase in $\Gamma_0$, the admissible range of $\lambda$ increases. Further, it is seen that as the model parameter $\lambda$ increases, both scalar spectral index and tensor-to-scalar ratio increases.  


\section{Conclusion}
Warm inflation emerged as an alternative theory for cold inflation which is consistent with standard big bang model. At the same time, modified theories of gravity have been developed to counter the shortcomings of GR. Both these stand in theories are important in order to unveil the mysteries of the universe. In this work, we have taken up warm inflation for study in the context of f(R,T) gravity in the strong dissipative regime $r > 1$. Further, we have chosen two potentials viz. Chaotic and Natural potential for study in the case of constant as well as variable dissipation coefficient $\Gamma$. The results are summarized below:
\begin{itemize}
    \item {\bf Chaotic potential with $\Gamma$ = constant} \\
    It is found that spectral index $n_s$ does not depend on the model parameter $\lambda$ whereas tensor-to-scalar ratio $R$ depends on the model parameter $\lambda$ and other parameters. For $n = 3$ and 4, spectral index and tensor-to-scalar ratio are consistent with the Planck 2018 bound for $N = 50/60/70$ e-folding. The allowed range of $f(R,T)$ model parameter is found to be $\lambda > 18$ when $n=3$ and $\lambda > 30$ when $n=4$. Other cases with $n = 2/3, 1, 2$ are ruled out since the scalar spectral index in this cases does not match Planck 2018 bound. 
    
    \item {\bf Chaotic potential with $\Gamma=\Gamma_0\frac{\tau^3}{\phi^{2}}$ } \\
    When we considered the variable form of $\Gamma$, it is found that only $n=2$ model gives result consistent with Planck 2018 data. The range of model parameter is found to be $145 < \lambda < 165$ for $\Lambda = 10^{-6}$, $152 < \lambda < 191$ for $\Lambda = 10^{-7}$ and $120 < \lambda < 145$ for $\Lambda = 10^{-8}$. It is seen that with the increase in model parameter $\lambda$ and with the decrease in the value of coupling constant $\Lambda$, tensor-to-scalar $R$ ratio decreases. If the  range of $R$ is further constrained, the correction from $f(R,T)$ gravity will help this model to remain consistent with the observational bound. 
    
    \item 
    {\bf Natural potential with $\Gamma$ = constant} \\
    In this case the inflationary scale $\mu$ is set at GUT scale and carried out the calculations. It is found that both the spectral index $n_s$ and tensor-to-scalar ratio $R$ depend on the model parameter $\lambda$ which means correction from $f(R,T)$ gravity has been induced in the theory. It is seen that both $n_s$ and $R$ remain within the Planck 2018 bound at $f \sim M_P$ scale. As we go on decreasing $f$ below Planck scale, spectral index and tensor-to-scalar ratio remain inside the Planck 2018 bound provided the model parameter space shifts to the higher ranges. This clearly indicates that within the framework of $f(R,T)$ gravity, the spontaneous symmetry breaking scale $f$ can be lowered below Planck scale and still the model will remain consistent with observational bounds. Hence $f(R,T)$ gravity solves the problem associated with Natural potential and makes it a reliable candidate for inflation at the sub Planckian scale.

    \item {\bf Natural potential with $\Gamma=\Gamma_0\frac{\tau^3}{\phi^{2}}$ } \\
    In this case the two energy scales $\mu$ and $f$ are set at sub Planckian scale. It is found that spectral index and tensor-to-scalar ratio depend on the model parameter $\lambda$ and are consistent with Planck 2018 data. The range of the model parameter is found to be $680 < \lambda < 780$ for $\Gamma_0 = 5 \times 10^{3}$, $680 < \lambda < 1150$ for $\Gamma_0 = 10^{4}$ and $680 < \lambda < 1150$ for $\Gamma_0 = 10^{5}$. Further, it is seen that as the model parameter $\lambda$ increases, both scalar spectral index and tensor-to-scalar ratio increases. So in this case lower values of model parameter is feasible for the model to be valid at sub Planckian scales.

\end{itemize}

In cold inflation, the correction from $f(R,T)$ gravity can't save Chaotic and Natural potential from rejection.\cite{r38} Now, in this work it is found that both the potentials are able to predict desirable results in the context of $f(R,T)$ gravity and warm inflation scenario. Further, the correction from $f(R,T)$ gravity is able to lower down the energy scales to the sub Planckian scales in the warm inflation which makes the potentials consistent with high energy particle physics. However, these results are specifically valid for the particluar form of $f(R,T)$ gravity used in this paper. For other forms of $f(R,T)$, results may vary. \\

Now as a possible extension of our work, one can study warm inflation for different other forms of ${f(R,T)}$ gravity which may provide interesting results. Besides this, one can use other rejected potentials in this framework and check whether they are saved by the correction term present in the ${f(R,T)}$ gravity. For a plethora of potential, one may check this reference.\cite{r47}

\end{document}